# Students taught by a first-time instructor using active learning teaching strategies outperform students taught by a highly-regarded traditional instructor

Colin S. Wallace[1]*, Edward E. Prather[2], John A. Milsom[3], Ken Johns[3], Srin Manne[3]

[1]Department of Physics and Astronomy, University of North Carolina at Chapel Hill, Chapel Hill, NC.

[2]Center for Astronomy Education (CAE), Steward Observatory, University of Arizona, Tucson, AZ.

[3]Department of Physics, University of Arizona, Tucson, AZ

*Correspondence to: cswphys@email.unc.edu

## Abstract

In this paper we put forth a model for physics course reform that uniquely uses proven, research-based active learning strategies to help students improve their physics knowledge and problem-solving skills.  In this study, we compared the exam performance of students in two sections of the same introductory physics course.  One section (the traditional section, $N = 258$) was taught by an instructor who is highly regarded for his lectures, but did not use any active learning teaching strategies.  The other section (the reformed section, $N = 217$) was taught by an instructor who had never before taught a physics class but who was trained in physics and astronomy education research and who did use active learning teaching strategies.  Students in the reformed section significantly outperformed students in the traditional section on common exam questions over the course of the semester, regardless of whether the question was conceptual or quantitative.  This reform effort has been successful at improving students' learning and significantly increasing the department's use of active learning strategies at the introductory level and beyond.

**Introduction**

This paper describes an effort to reform a physics course and to transform a department's attitudes toward active learning. In 2013, the University of Arizona became one of eight project sites funded by the American Association of Universities (www.aau.edu/aau-stem-project-sites) to redesign foundational STEM courses and to improve the attitudes of faculty toward active learning (Burd, *et al*., 2016). As part of this effort, the Physics Department conducted a study comparing two sections of the same introductory physics course taught using drastically different instructional approaches. One section was taught by an experienced and highly-regarded professor who used traditional lecture methods (the "traditional section"). The other was taught by a first-time physics instructor, versed in physics and astronomy education research, who used research-based active learning strategies (the "reformed section"). In order to achieve the desired change in the department's attitudes toward active learning, the reform section had to demonstrate improvement in students' quantitative problem-solving skills as well as their conceptual understandings. Since many of the active learning materials produced by physics education research (PER) focus on developing students' conceptual understandings (*e.g*., Hieggelke, Maloney, and Kanim 2012; Mazur 1997; McDermott and Shaffer 2008; O'Kuma, Maloney, and Hieggelke 2008), we had to adapt and develop active learning approaches to help improve students' abilities to solve quantitative problems. This paper describes all aspects of our reform effort: how we used active learning in novel ways to simultaneously promote quantitative problem-solving among 200+ students in a large lecture setting; how we compared the performance of students in the traditional and reformed sections; and how the results of this study instigated a cultural change in the Department of Physics.

Note that this reform effort was not initially envisioned as a research experiment. Consequently, we did not control for every possible variation in the two sections that might contribute to differences in student performance (e.g., differences in the homework assignments of the two sections, as well as some differences in topic pacing and depth of coverage). However, by the end of the semester, it was clear that the dramatic improvements in students' performance in the reformed section, over those of their peers in the traditional section, could be attributed to the in-class experiences afforded by the active-learning activities, rather than to some other variables that we did not control for.

This paper appears similar to the study of Deslauries, Schelew, and Weiman (2011). In that study, a first-time physics instructor served as a guest instructor for a single week in a section of a physics course. The first-time instructor made heavy use of active learning strategies while they taught. During the same week, an experienced instructor gave traditional lectures on the same content to students enrolled in a second section of the same course. Deslauries, Schelew, and Weiman found that students taught by the first instructor outperformed their peers taught by the experienced instructor on twelve multiple-choice questions designed to measure students' conceptual understandings.

Our study and course reform effort were significantly larger than and extend the work of Deslauries, Schelew, and Weiman (2011). We developed a semester-long curriculum, spanning the full range of topics commonly taught in introductory mechanics. Additionally, our reform effort included the development of a novel way to use active-learning strategies to foster quantitative problem-solving in the large-lecture environment. Finally, our assessments of students included both conceptual questions and quantitative problems, whereas Deslauries, Schelew, and Weiman only examined students' conceptual understandings. This paper expands the knowledge base on how to successfully implement research-based active learning instructional approaches to create sustainable course transformations.

**Reform effort**

Before describing the traditional and reformed sections in detail, we want to situate this study and reform effort in the campus-wide AAU-funded STEM education reform effort that was already underway at the University of Arizona. A major thrust this effort involved changing the culture and practices related to teaching in several STEM departments (Burd, *et al*., 2016). Research shows that cultural change practices that simply make innovative curriculum materials available to faculty, or that invoke "top-down" policies and mandates are ineffective. Successful reform efforts must endeavor to be both supportive of and compatible with the local culture, and utilize practices that are explicitly aligned with or work to change stakeholder's beliefs (Henderson, Beach, and Finkelstein, 2011). Furthermore, a long-term view must be part of the collective beliefs of the stakeholders, and so we set out to create a change that would persist well beyond the one semester of the study reported here. Consequently, this reform effort involved bringing multiple stakeholders together, including both instructors, the associate department head, the department's director of undergraduate studies, and other experts in PER to the table. We refer to this group as the "reform team."

The reform team held a series of meetings in the summer prior to the semester's instruction in order to establish a common set of course norms, policies, and goals for both the control and reformed sections. Throughout the semester, this group met weekly to develop and review particular active learning strategies, discuss the progress of both courses, ensure that both courses were covering the same material at roughly the same time (although there were some differences in topic pacing and coverage), and develop appropriate free-response exam questions that could assess students' physics content knowledge, reasoning abilities and quantitative problem-solving skills.

**Comparing and contrasting the traditional and reformed sections**

The course at the heart of this experiment was the first semester of introductory, calculus-based physics. 258 students enrolled in the traditional section and 217 enrolled in the reformed section. We made no effort to advertise the differences between the two sections while students were enrolling. The content of the course focuses on traditional physics topics such as kinematics, Newton's laws of motion, impulse and momentum, work and energy, rotational motion, gravitation, periodic motion, and fluids. The course is primarily taken by students pursuing degrees in the physical sciences and engineering.

There were no statistically significant differences between students in the traditional and reformed sections in terms of their SAT/ACT scores, scores on a math placement exam, Pell grant eligibility, and percentages of males versus females. There were also no statistically significant differences between the two sections in terms of the representation of different races and ethnicities among enrolled students. An independent samples *t*-test did reveal that students in the reformed section had a lower average high school GPA (3.58) than students in the traditional section (3.69) at a statistically significant level ($p = 0.008$) (McGrath, 2015).

Students from both the traditional and reformed sections attended a 170-minute laboratory session each week. We will not comment further on these labs, since they were identical between the traditional and reformed sections.

Students in the traditional section had three 50-minute lectures each week, whereas students in the reformed section only had two 50-minute lectures each week. The reformed section gave up a full 50-minute lecture each week so that students could attend a weekly small-group ($N \approx 25$) recitation section, which will be described in more detail in the next section. The presence of these recitations sections meant that while the reformed course had 50 fewer minutes in the lecture hall each week, the total amount of time students were expected to be in class was identical between the two sections. Since the exams for both sections occurred on the same day, both instructors had to make sure that they had covered the relevant content by that date – thus the reformed section had to provide instruction on new content with two, rather than three, lectures a week.

No "Studio" or "SCALE-UP" style rooms were available for either section. Instead, lectures for both sections were held in the same large lecture hall, with stadium-style seating and desks that were bolted to the ground. The lectures were held back-to-back in the morning.

We made sure that all graded elements of the course were weighted equally between the two sections. Homework accounted for 15% of their final course grade, laboratories accounted for 20%, the three midterms together constituted 40% of their final grade, and the final exam accounted for 25%. The common exams were all held in the evenings at the same time in order to eliminate the possibility that students from one section might divulge the content of the exam to students from the other section. Final letter grades were assigned to students from both sections based on the same scale. Both sections also used the same textbook (Young and Freedman 2012). All of these details were agreed upon by the reform team before the semester began.

**Active learning in the reformed section**

Each instructor created his own content for the lectures and assigned his own homework problems. Since the reformed section only met for lecture twice a week, the first-time instructor did not devote a substantial amount of class time to reproducing complex derivations from the textbook, nor did he introduce mathematical formalism (*e.g.*, vector calculus) at the start of the semester, instead opting to provide the relevant mathematics needed for the physics being studied on the day the topic was introduced (*e.g.*, the dot product of two vectors was introduced when students began studying work). Rather than focusing class time on derivations of equations, the reformed section's lectures focused on developing students' conceptual understandings, reasoning abilities, quantitative problem-solving skills, and fluencies with multiple-representations of the discipline. To accomplish this, the first time instructor made use of multiple research-validated and research-informed active-learning strategies, including Think-Pair-Share (aka "TPS" or "Peer Instruction"; Lyman 1981; Mazur 1997) and Ranking Tasks (RTs) (Hieggelke, Maloney, and Kanim 2012; O'Kuma, Maloney, and Hieggelke 2008). As noted above, these instructional strategies are typically used to improve students' conceptual understandings and reasoning abilities.

Before unpacking how the reformed section used active learning to also increase students' quantitative problem solving skills, consider how typical physics classes, including the traditional section, typically teach students how to problem solve. In a typical physics class, the instructor chooses a few problems that they use to model problem solving in front of students. Those students passively watch the instructor and take notes. Both the instructor and students hope that, through this process, students will learn how to interpret a problem, apply the relevant physics, develop a solution plan, and successfully execute the necessary calculations. Since students are not actively practicing these skills in class, both the instructor and students may not realize the degree to which students are struggling until after the homework or exams have been graded – and by then the course has moved on to new material.

Office hours are one place where faculty can gain an appreciation for students' struggles and provide students with the help they need to learn the material. However, since the students who attend are self-selected, this can easily misrepresent the primary difficulties of the class as a whole. Office hours are also an inefficient way to help a large number of students overcome whatever difficulties they are experiencing with the material. At the heart of our reform effort was the desire to bring the benefits of "office hours" to the large lecture classroom. Consequently, we re-conceptualized the use of "lecture time" during the reformed section. The active learning model we created was designed to promote an "office hours experience" for the entire class at once in order to help them integrate their developing conceptual understandings with their emerging quantitative problem-solving abilities.

During each class, students in the reformed section were presented with a problem and given a few minutes to attempt a solution while working in collaboration with their neighbors (see the Appendix for an example problem). The first-time instructor carefully chose problems that would elicit multiple and well-understood problem-solving and conceptual difficulties. While students worked toward a solution, the instructor would roam the room and answer questions. This brief initial problem-solving phase was critical since it afforded students the opportunity to interpret the problem, organize their ideas, and reason about how their physics knowledge might be applied to reach a solution for the problem. This initial phase helped students become aware of where along the solution pathway they "felt stuck" – an essential component of the in-class problem-solving curriculum.

The next phase of the curriculum used the questioning and voting strategy often referred to as Think-Pair-Share (TPS). In this second phase, students were asked to vote on multiple-choice questions designed to unpack their conceptual and reasoning difficulties associated with solving the particular quantitative problem on which they had just attempted a solution. What made our use of TPS unique was that the choices modelled incorrect mathematical expressions that highlighted fundamental errors and problem-solving difficulties common to our student population. The sequence of carefully crafted TPS questions focused students' thinking on specific steps along the problem solution pathway and it instigated peer conversations about the particular mathematical or physics applications known to be problematic for students (see the Appendix for example TPS questions). This novel use of mathematical expressions for choices to TPS questions forced students in class to unpack their mental models and think critically about quantitative problem-solving, rather than simply spend their class time hurriedly copying down the solution being created by the instructor. Our use of TPS allowed the first-time instructor, by himself (no teaching assistants were present in lecture), to simultaneously engage over 200 students in collaborative problem solving during each class, all without having to collect and grade any additional work, while also addressing all the content being presented in the traditional section. Students in the reformed section gained experience identifying and applying relevant physical principles, monitoring their own understandings, and actively focusing their mental efforts on difficult steps in the quantitative problem solving processes that physicists hold in high regard. For more details on how to effectively implement TPS for quantitative problem-solving, see Wallace (2020).

      The reformed section's weekly recitation sessions focused on further developing students' conceptual understandings and problem-solving skills, but did not introduce any new content. Each session had approximately 25 students and was led by a graduate teaching assistant. An undergraduate learning assistant (Otero, Finkelstein, McCray, and Pollock 2006) was also present to help facilitate student learning. During each recitation, students worked in

collaborative groups of three to four to solve a single quantitative problem. They also worked through a tutorial or ranking task activity inspired by curricula developed by physics and astronomy education researchers (Hieggelke, Maloney, and Kanim 2012; McDermott and Shaffer 2008; O'Kuma, Maloney, and Hieggelke 2008). Students received a nominal amount of participation credit, which was incorporated into their homework grade, for their recitation work. The TAs and LAs were trained by the first-time instructor during weekly meetings. During these training meetings, they reviewed the coming week's activities and engaged in a version of situated apprenticeship (Prather and Brissenden 2009) in which they modelled authentic student difficulties and practiced using Socratic-style questioning techniques to help each other overcome those difficulties.

**Findings**

The big question of this reform effort was whether the focus of class time on collaborative group quantitative problem solving, driven by TPS questions, combined with the implementation of the weekly recitation section, would be enough to help students develop greater conceptual understandings and significantly improve their quantitative problem-solving skills as compared to their peers in the traditional section. To answer this question, we compared the performance of students in the two sections on the four common exams they took over the course of the semester.

The exams were created by the reform team well in advance of the instruction students in either course received on the relevant physics. The exams used both conceptual and quantitative free-response questions. One-third of the questions on each exam were conceptual and the remaining two-thirds were quantitative. The majority of the conceptual questions were derived from questions used in and validated by PER, such as the passing twice speed comparison task

(Trowbridge and McDermott 1980), the work-energy and impulse-momentum tasks (Lawson and McDermott 1987; Pride, Vokos, and McDermott 1998), and the five blocks problem (Heron 2004). The quantitative questions were chosen by the reform team to have a level of difficulty comparable to questions used in previous versions of the course. This allowed us to align our assessment practices with the cultural expectations of the department, which was important for leveraging any successes of the reformed sections toward the larger culture-change efforts.

All exam questions were graded by graduate TAs assigned to the traditional and reformed sections. The instructors from both sections collaborated to develop grading rubrics for the TAs to follow, trained the TAs on how to apply the rubrics, and periodically spot checked graded exams to make sure that each question was being scored appropriately. The instructors also assigned a single question to a single TA, who then graded the responses of all the students in both sections to that question. This ensured that students in both sections were graded on the same scale and by the same person.

Since the classes differed with respect to the depth to which they covered certain topics, it is possible that students in one section may have done better on a particular exam item because they received more instruction on that topic and/or they saw a similar question during instruction. To investigate this possibility, one member of the reform team (not one of the section instructors) went through both sections' lecture slides, homework, and, in the case of the reformed section, recitation activities and compared each class's content to the exam questions. He flagged all items that could have been biased toward one class or the other and then shared these results with the rest of the reform team to discuss. After discussion, we decided to keep items that use standard representations of the discipline (e.g., representing vectors with arrows), present contexts frequently encountered in introductory physics classes and textbooks (e.g., constant acceleration and projectile motion problems), and require quantitative procedures that are always taught in introductory physics (e.g., using conservation principles to solve 2D inelastic collision problems). We removed items from our analysis whose representations, contexts, or quantitative procedures went beyond what is described in the previous sentence.

Table 1 shows students' average percent correct for both sections on every exam item, with biased items shaded. The listed uncertainties represent the standard deviation of the mean. Table 1 also shows, for each item, the difference ($\Delta$) between the average score of the reformed section and the traditional section. Positive values (which are italicized, bolded, and in blue) indicate that the reformed section had the higher average score, while negative values (which are bolded and in red) indicate that the traditional section had the higher average score. Only item 1 on exam 3 had a $\Delta$ indistinguishable from 0, so it is not bolded, italicized, or colored.

While in principle a question could have been biased in favor of either class, all the questions highlighted in Table 1 were biased in favor of the reformed class. This may be due to the fact that significant scaffolding and active engagement instruction is necessary for students to develop robust, physically-correct understandings that enable them to successfully answer these questions. We should note that the reformed section did spend more time unpacking the contexts, representations, and problem-solving procedures used on many of the flagged items. This was done intentionally. Many students in introductory physics are novices at the subject when they begin the course. To help them become more expert-like requires explicit mentoring and modelling of the abilities and practices that represent discipline expertise. These abilities and practices should be decided upon in advance and exemplified in the instruction and assessments provided to students (Wiggins and McTighe 2005).

For an example of an item that is biased, consider item 4 on exam 2. This item comes from the famous work-energy and impulse-momentum tasks designed by the University of Washington's PER group (Lawson and McDermott 1987; Pride, Vokos, and McDermott 1998). Students in the reformed section, unlike the traditional section, received explicit instruction on when to apply the work-energy theorem and when to apply the impulse-momentum theorem, and this instruction was reinforced in parts of two recitation activities.

Even if we ignore all items flagged as biased toward the reformed section, the reformed section had higher average scores on almost every question, with only two exceptions (items 1 and 4 on exam 1). The reformed class outperformed the traditional class on all other items on all exams, and in many cases the differences in scores were large (e.g., the difference on item 4 on the final exam was $18.6\% \pm 2.9\%$).

Figure 1 shows the average scores of each class on each exam. Note that we removed all items flagged as biased toward the reformed section when calculating these averages. The error bars represent the standard deviation of the mean.

Note that the two sections performed roughly the same on the first midterm, and both sections did quite well. This result is not surprising since the content of the first exam covered topics many students study in high school (kinematics and Newton's laws of motion). However, the second exam was significantly more difficult as it addressed work-energy and impulse-momentum principles, with which many students have less experience. On the second exam, students in the reformed section outperformed students from the traditional section, with scores that were approximately 5-14% greater than the scores of their peers in the traditional section. This result held regardless of the question type (conceptual or quantitative). This same trend is present on the third exam and the final exam. What is especially noteworthy about the final exam is the magnitude of the item differences. On some items, students from the reformed section outperformed their peers in the traditional section by almost 20%. Notice that the average scores on exams 2 and 3 for the reformed section were higher than those of the traditional section (by 8.5% and 7.5%, respectively). On the final exam, the difference between the sections was 13.2% in favor of the reformed section. Since the final exam was a cumulative exam, this result suggests that the effects of active learning may be cumulative: As students spend more time actively engaging with the material, they experience increased abilities to recall, apply, and synthesize what they have learned, even in contexts that are *prima facie* novel.

**Conclusions**

We conclude that students in the reformed section, taught by a first-time instructor using active learning teaching strategies, outperformed their peers in the traditional section, which was taught by a highly-regarded and very popular traditional instructor whose instruction relied primarily on lecture. This result is quite robust, since it is virtually insensitive to which exam items are included or excluded from the analysis. The results of this study strongly suggest that the use of active learning teaching strategies, including collaborative problem-solving in the large-lecture environment, is more effective at supporting student learning of physics – even when the course is taught by a first-time instructor.

The reform effort described in this paper was also successful in affecting a cultural change in the University of Arizona's Department of Physics. Since this experiment was run, the department has transitioned to making the recitation sections a permanent feature of all their introductory courses. They have also continued to develop, adapt, and adopt curricula and instructional techniques that are informed by the results of DBER in general and PER in particular. Numerous departmental stakeholders, including, but not limited to, members of the reform team, have successfully pushed to increase the use of active learning pedagogical techniques in other courses offered by the department. The recitation materials the department is using in the calculus-based mechanics course discussed in this paper and the materials in all of their other courses are all available at https://w3.physics.arizona.edu/undergrad/teaching-resources. The broader reform efforts of the department have been galvanized by our implementation of TPS for interactive problem solving, which has been shown to be easy to use and highly successful at increasing students' physics knowledge. We hope this reform model will inspire other instructors at other institutions to transform their courses to help their students.

## Acknowledgements

We thank Shelley McGrath for her work as the external evaluator for this project. This work was supported by the American Association of University's Undergraduate STEM Education Initiative.

|  |  | Exam 1 | Exam 2 | Exam 3 | Final Exam |
|---|---|---|---|---|---|
| Item 1 | Ref | 86.6±1.1 | 47.0±2.2 | 79.8±1.6 | 81.2±1.8 |
| Item 1 | Trad | 90.1±1.0 | 38.9±1.81 | 79.6±1.6 | 70.5±2.0 |
| Item 1 | Δ | *-3.6±1.4* | *8.1±2.8* | 0.1±2.3 | *10.7±2.6* |
| Item 2 | Ref | 75.6±1.8 | 43.8±2.5 | 50.0±2.1 | 43.9±2.2 |
| Item 2 | Trad | 70.4±1.7 | 30.0±2.27 | 42.5±2.1 | 33.1±1.8 |
| Item 2 | Δ | *5.3±2.4* | *13.9±3.4* | *7.5±3.0* | *10.8±2.8* |
| Item 3 | Ref | 95.5±0.9 | 53.5±2.2 | 49.6±1.9 | 85.2±1.5 |
| Item 3 | Trad | 92.3±1.3 | 46.3±2.33 | 40.9±1.9 | 66.0±2.1 |
| Item 3 | Δ | *3.3±1.6* | *7.2±3.2* | *8.6±2.7* | *19.2±2.6* |
| Item 4 | Ref | 53.7±2.3 | 49.2±2.3 | 69.8±1.6 | 87.5±1.7 |
| Item 4 | Trad | 62.4±2.2 | 41.0±2.0 | 57.9±1.8 | 68.9±2.4 |
| Item 4 | Δ | *-8.6±3.2* | *8.2±3.1* | *11.9±2.4* | *18.6±2.9* |
| Item 5 | Ref | 92.3±1.3 | 82.8±1.8 | 79.8±1.6 | 63.7±1.7 |
| Item 5 | Trad | 89.9±1.3 | 76.5±2.1 | 75.1±1.5 | 57.4±1.7 |
| Item 5 | Δ | *2.4±1.8* | *6.4±2.8* | *4.7±2.2* | *6.2±2.4* |
| Item 6 | Ref | 79.0±1.6 | 77.3±1.50 | 86.6±1.2 | 83.5±1.1 |
| Item 6 | Trad | 62.6±1.7 | 72.9±1.9 | 79.8±1.4 | 65.9±2.0 |
| Item 6 | Δ | *16.5±2.3* | *4.4±2.4* | *6.8±1.9* | *17.6±2.3* |
| Item 7 | Ref |  |  |  | 76.7±1.1 |
| Item 7 | Trad |  |  |  | 66.0±1.2 |
| Item 7 | Δ |  |  |  | *10.7±1.6* |
| Item 8 | Ref |  |  |  | 46.0±2.3 |
| Item 8 | Trad |  |  |  | 39.8±2.0 |
| Item 8 | Δ |  |  |  | *6.3±3.0* |
| Item 9 | Ref |  |  |  | 71.6±1.6 |
| Item 9 | Trad |  |  |  | 63.0±1.6 |
| Item 9 | Δ |  |  |  | *8.6±2.3* |

Table 1. The average score of students in the reformed and traditional sections on every exam item. The Δ represents the difference between the average for the reformed section and the average for the traditional section; positive values (which are also italicized and the color blue)

indicated the reformed section had a higher average while negative values (which are the color red) indicate the traditional section did better. Errors represent the standard deviation of the mean. Items in dark grey boxes were flagged as biased toward the reformed section.

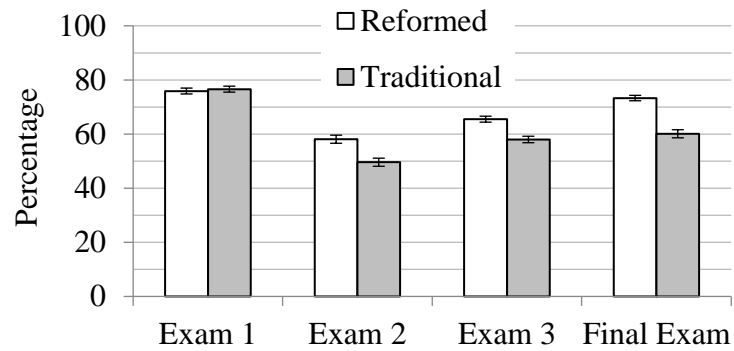

Figure 1. Average exam scores for both the traditional and reformed sections. Questions flagged as biased toward the reformed section were removed from the data set before calculating these averages. Error bars represent the standard deviation of the mean.

## Appendix

The following question is a typical example of the kinds of quantitative questions students in the reformed section were asked to solve during the lecture portion of class. Students were given the following prompt and diagram (which is based on exercise 8.41 in Young and Freedman, 13th edition) and asked to solve the problem while working collaboratively with their neighbors:

### Prompt and Diagram

A car of mass $m_c = 1500$ kg is traveling north through an intersection when it is hit by an SUV of mass $m_s = 2200$ kg traveling east. The two vehicles become locked together during the impact and slide together as one after the collision. The cars slide to a halt at a point 5.39 m east and 6.43 m north of the impact point. The coefficient of kinetic friction between the tires and the road is $\mu_k = 0.75$. What were the speed of the car ($v_c$) and the speed of the SUV ($v_s$) just before the impact?

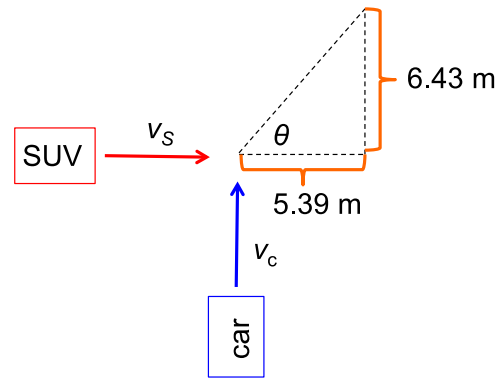

The first-time instructor would circulate around the room for several minutes while students worked, answering questions from individual groups. He would then bring the entire class back together and ask the following TPS questions in order to focus attention on and help students overcome particularly challenging aspects of the solution pathway.

### TPS Question 1

If $W_f$ is the work done by kinetic friction while the two connected cars slide across the ground, then which of the following is true?

A) $W_f = (½\, m_s v_s^2 + ½\, m_c v_c^2) - ½(m_s + m_c)(v_{s+c})^2$

B) $W_f = (½\, m_s v_s^2 + ½\, m_c v_c^2)$

C) $W_f = 0 - ½(m_s + m_c)(v_{s+c})^2$

D) $W_f = (½\, m_s v_s^2 + ½\, m_c v_c^2) + ½(m_s + m_c)(v_{S+c})^2$

*Commentary: Students must realize that they must apply both the work-energy theorem and the conservation of linear momentum in order to solve for the two unknown speeds. This question makes students think about how the work-energy theorem applies in the case of a completely inelastic collision.*

### TPS Question 2

What is the work done by kinetic friction?

A) $W_f = -\mu_k\, (m_s + m_c)\, g\, [(6.43\text{ m})^2 + (5.39\text{ m})^2]^{1/2}$

B) $W_f = -\mu_k\, (m_s + m_c)\, g\, [(6.43\text{ m})^2 + (5.39\text{ m})^2]$

C) $W_f = -\mu_k\, (m_s + m_c)\, g\, (6.43\text{ m} + 5.39\text{ m})$

D) $W_f = -\mu_k\, (m_s + m_c)\, g\, (6.43\text{ m})$

E) $W_f = -\mu_k\, (m_s + m_c)\, g\, (5.39\text{ m})$

*Commentary: This question helps students reason about the distance the two vehicles travel after the collision, which is necessary for calculating the work done by the non-conservative friction force. The answer choices for this question deliberately blend symbolic notation and specific numbers in order to facilitate students' abilities to efficiently reason through the five options.*

### TPS Question 3

Which of the following is the conservation of linear momentum applied to the x-components

of the momenta:

A) $m_s v_s + m_c v_c = (m_c + m_s) v_{s+c}$

B) $m_s v_s + m_c v_c = (m_c + m_s) v_{s+c} \tan(\theta)$

C) $m_s v_s = (m_c + m_s) v_{s+c} \cos(\theta)$

D) $m_c v_c = (m_c + m_s) v_{s+c} \sin(\theta)$

E) more than one of the above

*Commentary: This question requires students to think about the conservation of linear momentum and how it can be used to relate the speed of the vehicles after they collide to their pre-collision speeds. This question also forces students to reason about which trigonometric function they need to use, which is a frequent source of confusion among introductory physics students.*

## TPS Question 4

Which of the following is the conservation of linear momentum applied to the *y*-components of the momenta:

A) $m_s v_s + m_c v_c = (m_c + m_s) v_{s+c}$

B) $m_s v_s + m_c v_c = (m_c + m_s) v_{s+c} \tan(\theta)$

C) $m_s v_s = (m_c + m_s) v_{s+c} \cos(\theta)$

D) $m_c v_c = (m_c + m_s) v_{s+c} \sin(\theta)$

E) more than one of the above

*Commentary: This question is the same as TPS Question 3, except for the y-components.*

## TPS Question 5

What is cos(θ)?

A) 5.39 / 6.43

B) 6.43 / 5.39

C) 5.39 / (5.39 + 6.43)

D) 6.43 / (5.39 + 6.43)

E) 5.39 / (5.39² + 6.43²)^(1/2)

*Commentary: Students sometimes need to be reminded of the definition of trigonometric quantities in terms of the sides of a right triangle.*

After this final question, students were given a few more minutes to calculate a solution. The first-time instructor then asked for the class to shout out the initial speed of the car and the SUV before revealing the solution:

### Solution

First apply the work-energy theorem:

$$v_{S+C}^2 = 2m_k g \sqrt{(6.43 \text{ m})^2 + (5.39 \text{ m})^2} = 123 \text{ m}^2/\text{s}^2$$
$$v_{S+C} = 11.1 \text{ m/s}$$

Next apply the conservation of momentum:

$$v_c = \frac{(m_c + m_s)v_{s+c}\sin(q)}{m_c} = \frac{(1500 \text{ kg} + 2200 \text{ kg})(11.1 \text{ m/s})}{1500 \text{ kg}} \frac{6.43}{\sqrt{6.43^2 + 5.39^2}} = 21.0 \text{ m/s}$$

$$v_s = \frac{(m_c + m_s)v_{s+c}\cos(q)}{m_s} = \frac{(1500 \text{ kg} + 2200 \text{ kg})(11.1 \text{ m/s})}{1500 \text{ kg}} \frac{5.39}{\sqrt{6.43^2 + 5.39^2}} = 12.0 \text{ m/s}$$